\documentclass[reqno,11pt]{amsart}
\usepackage{multind}
\usepackage{hyperref}
\usepackage{graphicx}
\usepackage{amscd}
\usepackage[mathscr]{eucal}
\numberwithin{equation}{section}
\allowdisplaybreaks[1]

\newcommand{\SetFigFont}[3]{}

\title[A New Formulation of Quantum Field Theory]{
A Formulation of Quantum Field Theory \\
Realizing a Sea of Interacting Dirac Particles}
\author[F.\ Finster]{Felix Finster \\ \\ November 2009 / July 2010}
\thanks{Supported in part by the Deutsche Forschungsgemeinschaft.}
\address{Fakult\"at f\"ur Mathematik \\ Universit\"at Regensburg \\ D-93040 Regensburg \\ Germany}
\email{Felix.Finster@mathematik.uni-regensburg.de}

\newtheorem{Def}{Definition}[section]

\newcommand{\Thanks}{\vspace*{.6em} \noindent \thanks}
\newcommand{\beq}{\begin{equation}}
\newcommand{\eeq}{\end{equation}}
\newcommand{\Proof}{\begin{proof}}
\newcommand{\QED}{\end{proof} \noindent}

\newcommand{\bra}{\,<\!\!}
\newcommand{\ket}{\!\!>\,}

\newcommand{\Pdd}{\mbox{$\partial$ \hspace{-1.2 em} $/$}}
\newcommand{\slsh}{\mbox{ \hspace{-1.13 em} $/$}}
\newcommand{\Aslsh}{\mbox{ $\!\!A$ \hspace{-1.2 em} $/$}}

\renewcommand{\L}{{\mathcal{L}}}
\newcommand{\Sact}{{\mathcal{S}}}
\newcommand{\T}{{\mathcal{T}}}
\newcommand{\B}{{\mathscr{B}}}

\begin{document}

\begin{abstract}
In this survey article, we explain a few ideas behind the fermionic projector approach
and summarize recent results which clarify the connection to quantum field theory.
The fermionic projector is introduced, which describes the physical system by a collection of
Dirac states, including the states of the Dirac sea.
Formulating the interaction by an action principle for the
fermionic projector, we obtain a consistent description of interacting quantum fields
which reproduces the results of perturbative quantum field theory.
We find a new mechanism for the generation of boson masses and obtain small
corrections to the field equations which violate causality.
\end{abstract}

\maketitle

\section{Introduction and Motivation} \label{secintro}
In order to give the negative-energy solutions of the Dirac equation a meaningful
physical interpretation, Dirac proposed that in the vacuum all states of negative energy should
be occupied by particles forming the so-called {\em{Dirac sea}}~\cite{dirac2, dirac3}.
His idea was that the homogeneous and isotropic Dirac sea configuration of the vacuum
should not be accessible to measurements, but deviations from this uniform configuration
should be observable. Thus particles are described by occupying additional states having
positive energy, whereas ``holes'' in the Dirac sea can be observed as anti-particles.
Moreover, Dirac noticed in~\cite{dirac3} that deviations from the uniform
sea configuration may also be caused by the interaction with an electromagnetic field. In order
to analyze this effect, he first considered a formal sum over all vacuum sea states
\beq \label{seasum}
R(t, \vec{x}; t', \vec{x}') = \sum_{l \text{ occupied}} \Psi_l(t, \vec{x}) \:\overline{\Psi_l(t', \vec{x}')} \:.
\eeq
He found that this sum diverges if the space-time point~$(t, \vec{x})$ lies on the
light cone centered at~$(t', \vec{x}')$ (i.e.\ if~$(t-t')^2 = |\vec{x} - \vec{x}'|^2$).
Next, he inserted an electromagnetic potential into the Dirac equation,
\[ \big( i \Pdd + e \Aslsh(t, \vec{x}) - m \big) \Psi_l(t, \vec{x}) = 0 \:. \]
He proceeded by decomposing the resulting sum~\eqref{seasum} as
\beq \label{Rab}
R = R_a + R_b \:,
\eeq
where~$R_a$ is again singular on the light cone, whereas~$R_b$ is a regular function.
The dependence of~$R_a$ and~$R_b$ on the electromagnetic potential
can be interpreted as describing a ``polarization of the Dirac sea''
caused by the non-uniform motion of the sea particles in the electromagnetic field.

When setting up an interacting theory, one faces the problem that
the total charge density of the sea states is given by the divergent expression
\[ \sum_{l \text{ occupied}} e\: \overline{\Psi_l(t, \vec{x})} \gamma^0 \Psi_l(t, \vec{x}) \:. \]
Thus the Dirac sea has an infinite charge density,
making it impossible to couple it to a Maxwell field.
Similarly, the Dirac sea has an infinite negative energy density, leading to divergences in
Einstein's equations. Thus before formulating the field equations, one must
get rid of the infinite contribution of the Dirac sea to the current and the energy-momentum
tensor.

In the standard perturbative description of quantum field theory (QFT), this is accomplished
by subtracting infinite counter terms
(for a more detailed discussion also in connection to renormalization see Section~\ref{secvac} below).
Then in the resulting theory, the Dirac sea is no longer apparent. Therefore, it is a common
view that the Dirac sea is merely a historical relic which is no longer needed in
modern QFT. However, this view is too simple because
removing the Dirac sea by infinite counter terms entails conceptual problems.
The basic shortcoming can already be understood from the representation~\eqref{Rab} of the Dirac
sea in an electromagnetic field.
 Since the singular term~$R_a$ involves~$\Aslsh$,
the counter term needed to compensate the infinite charge density of the Dirac sea must
depend on the electromagnetic potential. But then it is no longer clear how precisely this counter term
is to be chosen. In particular, should the counter term include~$R_b$,
or should~$R_b$ not be compensated and instead enter the Maxwell equations?
More generally, in a given external field, the counter terms involve the background field, giving
a lot of freedom in choosing the counter terms. In curved space-time, the situation is even more
problematic because the counter terms depend on the choice of coordinates.
Taking the resulting arbitrariness seriously,
one concludes that the procedure of subtracting infinite charge or energy densities is not a fully
convincing concept. Similarly, infinite counter terms are also needed in order to treat the
divergences of the Feynman loop diagrams.
Dirac himself was uneasy about these infinities, as he
expressed later in his life in a lecture on quantum electrodynamics~\cite[Chapter~2]{dirac4}:
\begin{quote}
``I must say that I am very dissatisfied with the situation, because this so-called good theory does involve
neglecting infinities which appear in its equations \ldots in an arbitrary way.
This is not sensible mathematics. Sensible mathematics involves neglecting a quantity when it turns
out to be small -- not neglecting it just because it is infinitely great and you do not want it!''
\end{quote}

The dissatisfaction about the treatment of the Dirac sea in perturbative QFT
was my original motivation for trying to set up a QFT
where the Dirac sea is not handled by infinite counter terms, but where the states of the Dirac sea are
treated on the same footing as the particle states all the way, thus making Dirac's idea of
a ``sea of interacting particles'' manifest.
The key step for realizing this program is to describe the interaction
by a new type of action principle, which has the desirable property that the divergent terms
in~\eqref{seasum} drop out of the equations, making it unnecessary to subtract any counter terms.
This action principle was first introduced in~\cite{PFP}.
More recently, in~\cite{sector} it was analyzed in detail for a system of
Dirac seas in the simplest possible configuration referred to as a single sector.
Furthermore, the connection to entanglement and second quantization was clarified in~\cite{entangle}.
Putting these results together, we obtain a consistent formulation of QFT
which is satisfying conceptually and reproduces the results of perturbative QFT.
Moreover, our approach gives surprising results which go beyond standard QFT,
like a mechanism for the generation of boson masses and small corrections to the field equations
which violate causality. The aim of the present paper is to explain a few ideas behind the
fermionic projector approach and to review the present status of the theory.

\section{Perturbative Quantum Field Theory and its Shortcomings} \label{secvac}
Let us revisit the divergences in~\eqref{seasum} in the context of modern QFT.
Historically, Dirac's considerations were continued by Heisenberg~\cite{heisenberg2}, who analyzed
the singularities of~$R_a$ in more detail and used physical arguments involving
conservation laws and the requirement of gauge
invariance to deduce a canonical form of the counter terms in Minkowski space.
This result was then taken up by Uehling and Serber~\cite{uehling, serber} to deduce corrections
to the Maxwell equations which are now known as the one-loop vacuum polarization.
A more systematic analysis became possible by {\em{covariant perturbation
theory}} as developed following the pioneering work of
Schwinger, Feynman and Dyson (see for example~\cite{schwinger, feynman, dyson2}).
In the resulting formulation of the interaction in terms of Feynman diagrams, one can
compute the loop corrections and the $S$-matrix of a scattering process,
in excellent agreement with experiments.
Moreover, the procedure of subtracting infinite counter terms was replaced by the
{\em{renormalization program}}, which can be outlined as follows
(for details cf.~\cite{peskin+schroeder} or~\cite{collins}): In order to get rid of the divergences
of the Feynman diagrams, one first regularizes the theory. Then one shows that the regularization
can be removed if at the same time the coupling constants and masses in the theory are suitably
rescaled. Typically, the coupling constants and the masses diverge as the ultraviolet
regularization is removed, but in such a way that the effective theory obtained in the limit
has finite effective coupling constants and finite effective masses.
The renormalization program is carried out order by order in perturbation theory.
Clearly, the procedure is not unique as there is a lot of freedom in choosing the
regularization. A theory is called {\em{renormalizable}} if this freedom can be described
to all orders in perturbation theory by a finite number of empirical constants.

Despite its overwhelming success, the present formulation of QFT suffers from
serious shortcomings. A major technical problem is that, despite considerable effort (see for
example~\cite{glimm+jaffe}), one has not succeeded in rigorously constructing an interacting
QFT in Minkowski space. In particular, the renormalized perturbation series of quantum electrodynamics
makes sense only as a formal power expansion in the coupling constant.
A more conceptual difficulty is that the covariant perturbation expansion makes statements only
on the scattering matrix. This makes it possible to compute the asymptotic in- and out-states
in a scattering process. But it remains unclear what the quantum field is at intermediate times,
while the interaction takes place. Moreover, one needs free asymptotic states
to begin with. But under realistic conditions, the system interacts at all times, so that
there are no asymptotic states. What does the quantum field mean in this situation?
For example, if one tries to formulate the theory in a fixed time-dependent background field,
then there are no plane-wave solutions to perturb from, so that standard perturbation theory fails.
If one tries to include gravity, the equivalence principle demands that the
theory should be covariant under general coordinate transformations.
But the notion of free states distinguishes specific coordinate systems, in which the free states are
represented by plane waves.
A related difficulty is entailed in the notion of the ``Feynman propagator'', defined by the conditions
that positive and negative frequencies should propagate to the future and past, respectively.
Again the notion ``frequency'' refers to an observer, explaining why Feynman's frequency conditions
are not invariant under general coordinate transformations.
To summarize, present QFT involves serious conceptual difficulties if one
wants to go beyond the computation of the scattering matrix and tries to understand
the dynamics of the quantum field at intermediate times or considers systems which for large times
do not go over to a free field theory in Minkowski space.

In order to understand these conceptual difficulties in more detail, it is a good starting point to
disregard the divergences caused by the interaction and to consider
{\em{free quantum fields in an external field}}.
In this considerably simpler setting, there are several approaches to construct
quantum fields, as we now outline.
Historically, quantum fields in an external field were first analyzed in the Fock space formalism.
Klaus and Scharf~\cite{klaus+scharf1, klaus+scharf2} considered
the Fock representation of the electron-positron field in the presence of a static external field.
They noticed that the Hamiltonian needs to be regularized by suitable counter terms which
depend on the external field. Thus the simple method of the renormalization program of removing the
regularization while adjusting the bare masses and coupling constants no longer works.
Similar to the explanation in Section~\ref{secintro}, one needs to subtract infinite counter terms
which necessarily involve the external field. Klaus and Scharf also realized that the Fock
representation in the external field is in general inequivalent to the standard Fock representation
of free fields in Minkowski space (see also~\cite{nenciu+scharf, klaus}).
This result shows that a perturbation expansion about the standard Fock vacuum necessarily
fails.

In the time-dependent setting, Fierz and Scharf~\cite{fierz+scharf} proposed
that the Fock representation should be adapted to the external field as measured by a local observer.
Then the Fock representation becomes time and observer dependent.
This implies that the distinction between particles and anti-particles no longer has an invariant meaning,
but it depends on the choice of an observer. In this formulation, the usual particle interpretation of
quantum states only makes sense for the in- and outgoing scattering states, but it has no invariant meaning
for intermediate times. For a related approach which allows for the construction of quantum fields in the
presence of an external magnetic field see~\cite{merkl}.
In all the above approaches, the Dirac sea leads to divergences, which must be treated by
an ultraviolet regularization and suitable counter terms.

As an alternative to working with Fock spaces, one can use the 
so-called {\em{point splitting renormalization method}},
which is particularly useful for renormalizing the expectation value of the energy-momentum
tensor~\cite{christensen}. Similar to the above procedure of Dirac and Heisenberg for treating
the charge density of the Dirac sea, the idea is to replace a function of one variable~$T(x)$ by
a two-point distribution~$T(x,y)$, and to take the limit~$y \rightarrow x$ after subtracting
suitable singular distributions which take the role of counter terms.
Analyzing the singular structure of the counter terms leads to the so-called {\em{Hadamard condition}}
(see for example~\cite{fulling+sweeny+wald}).
Reformulating the Hadamard condition for the two-point function as a
local spectral condition for the wave front set~\cite{radzikowski}
turns out to be very useful for the axiomatic formulation of free quantum fields in curved space-time.
As in the Fock space formalism, in the point splitting approach the particle interpretation depends
on the observer. This is reflected mathematically by the fact that the Hadamard condition
specifies the two-point distribution only up to smooth contributions.
For a good introduction to free quantum fields in curved space-time we refer to the
recent book~\cite{baer+fredenhagen}.

We again point out that in all the above papers on quantum fields in an external field or in curved
space-time, only free fields are considered. The theories are not set up in a way where it would be
clear how to describe an additional interaction in terms of Feynman diagrams.
Thus it is fair to say that the formulation of a background independent interacting perturbative
QFT is an open and apparently very difficult problem.
All the methods so far suffer from the conceptual difficulty that to avoid divergences,
one must introduce infinite counter terms ad-hoc.

\section{An Action Principle for the Fermionic Projector in Space-Time}
In order to introduce the fermionic projector approach, we now define
our action principle on a formal level (for the analytic justification 
and more details see~\cite[Chapter~2]{sector}).
Similar to~\eqref{seasum}, we describe our fermion system for any points~$x$ and~$y$
in Minkowski space by the so-called {\em{kernel of the fermionic projector}}
\beq \label{Pkernel}
P(x,y) = - \!\!\!\sum_{l \text{ occupied}} \Psi_l(x) \:\overline{\Psi_l(y)} \:,
\eeq
where by the occupied states we mean the sea states except for the anti-particle states
plus the particle states. For any~$x$ and~$y$, we introduce the
{\em{closed chain}}~$A_{xy}$ by
\beq \label{cchain}
A_{xy} = P(x,y)\, P(y,x)\:.
\eeq
It is a $4 \times 4$-matrix which can be considered as a linear operator on the Dirac wave functions
at~$x$. For such a linear operator~$A$ we define the {\em{spectral weight}} $|A|$ by
\[ |A| = \sum_{i=1}^4 |\lambda_i|\:, \]
where~$\lambda_1, \ldots, \lambda_4$ are the eigenvalues of~$A$ counted with algebraic
multiplicities. We define the {\em{Lagrangian}} $\L$ by
\beq \label{Ldef}
\L_{xy}[P] = |A_{xy}^2| - \frac{1}{4}\: |A_{xy}|^2 \:.
\eeq
Integrating over space-time, we introduce the functionals
\beq \label{STdef} \Sact[P] \;=\; \iint \L_{xy}[P] \:d^4 x\: d^4y \qquad \text{and} \qquad
\T[P] \;=\; \iint |A_{xy}|^2 \:d^4 x\: d^4y\:.
\eeq
Our action principle is to
\beq \label{actprinciple}
\text{minimize $\Sact$ for fixed~$\T$}\:,
\eeq
under variations of the wave functions~$\Psi_l$ which preserve the normalization
with respect to the space-time inner product
\beq \label{stip}
\bra \Psi | \Phi \ket = \int \overline{\Psi(x)} \Phi(x)\: d^4x\:.
\eeq

The action principle~\eqref{actprinciple} is the result of many thoughts and extensive calculations
carried out over several years. The considerations which eventually led to this action principle
are summarized in~\cite[Chapter~5]{PFP}. Here we only make a few comments.
We first note that the factor~$1/4$ in~\eqref{Ldef} is merely a convention, as the value of
this factor can be arbitrarily changed by adding to~$\Sact$ a multiple of the constraint~$\T$.
Our convention has the advantage that for the systems under consideration here, the Lagrange multiplier
of the constraint vanishes, making it possible to disregard the constraint in the following discussion.
Next, we point out that taking the absolute value of an eigenvalue of the closed chain
is a non-linear (and not even analytic) operation, so that our Lagrangian is not quadratic.
As a consequence, the corresponding Euler-Lagrange equations are {\em{nonlinear}}. Our Lagrangian has
the property that it vanishes if~$A$ is a multiple of the identity matrix. Furthermore, it vanishes if
the eigenvalues of~$A$
form a complex conjugate pair. These properties are responsible for the fact that the
singularities on the light cone discussed in the introduction drop out of the Euler-Lagrange
equations. Moreover, it is worth noting that the action involves only the fermionic wave
functions, but {\em{no bosonic fields}} appear at this stage. The interaction may be interpreted
as a direct particle-particle interaction of all the fermions, taking into account the sea states.
We finally emphasize that our action involves neither coupling constants nor any other
free parameters.

Clearly, our setting is very different from the conventional formulation of physics.
We have neither a fermionic Fock space nor any bosonic fields. Although the expression~\eqref{Pkernel}
resembles the two-point function, the $n$-point functions are not defined in our setting.
More generally, it seems inappropriate and might even be confusing to use notions from QFT,
which have no direct correspondence here. Thus one should be willing to accept that
we are in a new mathematical framework where we describe
the physical system on the fundamental level by the fermionic projector with kernel~\eqref{Pkernel}.
The connection to QFT is not obvious at this stage, but will be
established in what follows.

We finally remark that our approach of working with a nonlinear functional
on the fermionic states has some similarity to the ``non-linear spinor theory'' by Heisenberg
et al~\cite{heisenberg}, which was controversially discussed in the 1950s, but
did not get much attention after the invention of renormalization.
We point out that our action~\eqref{actprinciple} is
completely different from the equation~$\Pdd \Psi \pm l^2 \gamma^5 \gamma^j \Psi\: (\overline{\Psi} \gamma_j \gamma^5 \Psi) =0$ considered in~\cite{heisenberg}. Thus there does not seem to
be a connection between these approaches.

\section{Intrinsic Formulation in a Discrete Space-Time} \label{secdst}
Our action principle has the nice feature that it does not involve the differentiable,
topological or causal structure of the underlying Minkowski space. This makes it possible
to drop these structures, and to formulate our action principle intrinsically in a discrete space-time.
To this end, we simply replace Minkowski space by a finite point set~$M$.
To every space-time point we associate the {\em{spinor space}} as a four-dimensional
complex vector space endowed with an inner product of signature~$(2,2)$,
again denoted by~$\overline{\Psi} \Phi$.
A {\em{wave function}}~$\Psi$ is defined as a function which maps
every space-time point~$x \in M$ to a vector~$\Psi(x)$ in the corresponding spinor space.
For a (suitably orthonormalized) finite family of wave functions~$\Psi_1, \ldots, \Psi_f$ we
then define the kernel of the fermionic projector in analogy to~\eqref{Pkernel} by
\[ P(x,y) = -\sum_{l=1}^f \Psi_l(x) \overline{\Psi_l(y)} \:. \]
Now the action principle can be introduced again by~\eqref{STdef}--\eqref{stip}
if we only replace the space-time integrals by sums over~$M$.

The formulation in discrete space-time is a possible approach for physics on the Planck scale.
The basic idea is that the causal and metric structure should be induced
on the space-time points by the fermionic projector as a consequence of a spontaneous symmetry
breaking effect. In non-technical terms, this {\em{structure formation}} can be understood by a
self-organization of the wave functions as described by our action principle.
More specifically, a discrete notion of causality is introduced as follows:
\begin{Def} {\bf{(causal structure)}} \label{defcausal}
Two space-time points~$x,y \in M$ are called {\bf{timelike}} separated if
the spectrum of the product~$P(x,y) P(y,x)$ is real. Likewise, the points are
{\bf{spacelike}} separated if the spectrum of~$P(x,y) P(y,x)$ forms two complex
conjugate pairs having the same absolute value.
\end{Def} \noindent
We refer the reader interested in the spontaneous structure formation and the connection
between discrete and continuum space-times to the
survey paper~\cite{lrev} and the references therein. The only point of relevance for what
follows is that in the discrete formulation, our action principle is finite and minimizers exist.
Thus there is a fundamental setting where the physical equations
are intrinsically defined and have regular solutions without any divergences.

\section{Bosonic Currents Arising from a Sea of Interacting Dirac Particles}
In preparation for analyzing our action principle, we need a systematic method
for describing the kernel of the fermionic projector in position space.
In the vacuum, the formal sum in~\eqref{Pkernel} is made precise as the
Fourier integral of a distribution supported on the lower mass shell,
\beq \label{Psea}
P^\text{sea}(x,y) \;=\; \int \frac{d^4k}{(2 \pi)^4}\: (k\slsh+m)\:
\delta(k^2-m^2)\: \Theta(-k^0)\: e^{-ik(x-y)}
\eeq
(where~$\Theta$ is the Heaviside function). In order to introduce particles and anti-particles,
one occupies (suitably normalized) positive-energy states or removes states of the sea,
\beq \label{particles}
P(x,y) = P^\text{sea}(x,y)
-\frac{1}{2 \pi} \sum_{k=1}^{n_f} \Psi_k(x) \overline{\Psi_k(y)}
+\frac{1}{2 \pi} \sum_{l=1}^{n_a} \Phi_l(x)  \overline{\Phi_l(y)} \:.
\eeq

Next we want to modify the physical system so as to describe a general interaction. To this end, it is useful
to regard~$P(x,y)$ as the integral kernel of an operator~$P$ on the wave functions, i.e.
\[ (P \Psi)(x) := \int P(x,y)\: \Psi(y)\: d^4y \:. \]
Since we want to preserve the normalization of the fermionic states with respect to the
inner product~\eqref{stip}, the interacting fermionic projector~$\tilde{P}$ can be obtained from the
vacuum fermionic projector~$P$ by the transformation
\[ \tilde{P} = U P U^{-1} \]
with an operator~$U$ which is unitary with respect to the inner product~\eqref{stip}. The calculation
\[ 0 = U (i \Pdd - m) P U^{-1} = U (i \Pdd - m) U^{-1} \tilde{P} \]
shows that~$\tilde{P}$ is a solution of the Dirac equation
\[ (i \Pdd + \B - m) \tilde{P} = 0 \qquad \text{where} \qquad
\B := i U \Pdd U^{-1} - i \Pdd \:. \]
This consideration shows that we can describe a general interaction by a potential~$\B$ in the
Dirac equation, provided that~$\B$ is an operator of a sufficiently general form.
It can be a multiplication or differential operator, but it could even be a nonlocal operator.
The usual bosonic potentials correspond to special choices of~$\B$.
This point of view is helpful because then the bosonic potentials no longer need to be considered
as fundamental physical objects. They merely become a technical device for describing specific
variations of the Dirac sea.

In order to clarify the structure of~$\tilde{P}$ near the light cone, one
performs the so-called {\em{causal perturbation expansion}} and
the {\em{light-cone expansion}}. For convenience omitting
the tilde, one gets in analogy to~\eqref{Rab} a decomposition of the form
\beq \label{Pdecomp}
P^\text{sea}(x,y) = P^{\text{sing}}(x,y) + P^{\text{reg}}(x,y) \:,
\eeq
where~$P^{\text{sing}}(x,y)$ is a distribution which is singular on the light cone and can be
expressed explicitly by a series of terms involving
line integrals of~$\B$ and its partial derivatives along the line segment~$\overline{xy}$.
The contribution~$P^{\text{reg}}$, on the other hand, is a
smooth function which is noncausal in the sense that it depends on the global behavior of~$\B$
in space-time. It can be decomposed further into so-called low-energy and
high-energy contributions which have a different internal structure.

For simplicity, we here omit all details and only mention two points which
are important for the physical understanding. First, one should keep in mind that the
distribution~$P^\text{sea}$ as defined by the causal perturbation expansion distinguishes
a unique reference state, even if~$\B$ is time dependent. Thus the decomposition~\eqref{particles}
yields a globally defined picture of particles and anti-particles, independent of a local observer.
Second, it is crucial for the following constructions that the line integrals appearing
in~$P^\text{sing}$ also involve partial derivatives of~$\B$. In the case when~$\B=\Aslsh$ is an
electromagnetic potential (or similarly a general gauge field), one finds that~$P^\text{sing}$ involves
the electromagnetic field tensor and the electromagnetic current. More specifically, the contribution
to~$P^\text{sing}$ involving the electromagnetic current takes the form
\beq \label{current}
-\frac{e}{16 \pi^3} \int_0^1 (\alpha-\alpha^2) \gamma_k \, (\partial^k_{\;\:l} A^l - \Box A^k) \big|_{\alpha y + (1-\alpha x)}
\;\lim_{\varepsilon \searrow 0} \log \Big( (y-x)^2 + i \varepsilon \:(y^0-x^0) \Big) \:.
\eeq
The appearance of this contribution to the fermionic projector can be understood similar to
the ``polarization of the Dirac sea'' mentioned in the introduction as being a result of
the non-uniform motion of the sea particles in the electromagnetic field.
This contribution influences the closed chain~\eqref{cchain} and thus has an effect
on our action principle~\eqref{actprinciple}. In this way, the electromagnetic current also enters
the corresponding Euler-Lagrange equations.
In general terms, one can say that in our formulation, the bosonic currents arise in
the physical equations only as a consequence of the collective dynamics of the particles of the Dirac sea.

\section{The Continuum Limit, the Field Equations} \label{secfield}
We now outline the method for analyzing our action principle
for the fermionic projector~\eqref{Pdecomp}. Since~$P^\text{sing}$ is a distribution which
is singular on the light cone, the pointwise product~$P(x,y) \,P(y,x)$ is ill-defined.
Thus in order to make mathematical sense of the Euler-Lagrange equations corresponding to our
action principle, we need to introduce an ultraviolet regularization.
Such a regularization is not a conceptual problem because
the setting in discrete space-time in Section~\ref{secdst} can be regarded as a special
regularization. Thus in our approach, a specific, albeit unknown regularization should have
a fundamental significance. Fortunately, the details of this regularization are not needed for
our analysis. Namely, for a general class of regularizations of the vacuum Dirac sea
(for details see~\cite[Chapter~3]{sector} or~\cite[Chapter~4]{PFP}),
the Euler-Lagrange equations have a well-defined asymptotic behavior when the regularization is removed.
In this limit, the Euler-Lagrange equations give rise to differential equations involving the
particle and anti-particle wave functions as well as the bosonic potentials and currents, whereas the
Dirac sea disappears. This construction is subsumed under the notion {\em{continuum limit}}.

In the recent paper~\cite{sector}, the continuum limit was analyzed in detail for
systems which in the vacuum are described in generalization of~\eqref{Psea}
by a sum of Dirac seas,
\beq \label{seavac}
P^\text{sea}(x,y) \;=\; \sum_{\beta=1}^g
\int \frac{d^4k}{(2 \pi)^4}\: (k\slsh+m_\beta)\:
\delta(k^2-m_\beta^2)\: \Theta(-k^0)\: e^{-ik(x-y)} \:.
\eeq
Such a configuration is referred to as a {\em{single sector}}.
The parameter~$g$ can be interpreted as the number of generations of
elementary particles. It turns out that in the case~$g=1$ of one Dirac sea,
the continuum limit gives equations which are only satisfied in the vacuum,
in simple terms because the logarithm in current terms like~\eqref{current}
causes problems.
In order to get non-trivial differential equations, one must assume that
there are exactly {\em{three generations}} of elementary particles. In this case,
the logarithms in the current terms of the three Dirac seas can compensate each other,
as is made precise by a uniquely determined so-called local axial transformation.
Analyzing the possible operators~$\B$ in the corresponding Dirac equation in an
exhaustive way (including differential and nonlocal operators), one finds that the
dynamics is described completely by an {\em{axial potential}} $A_\text{a}$
coupled to the Dirac spinors. We thus obtain the coupled system
\beq \label{daeq}
(i \Pdd + \gamma^5 \Aslsh_\text{\rm{a}} - m) \Psi = 0 \:,\qquad
C_0 \,j^k_\text{\rm{a}} - C_2\, A^k_\text{\rm{a}} = 12 \pi^2\, J^k_\text{\rm{a}}\:,
\eeq
where~$j_\text{\rm{a}}$ and~$J_\text{\rm{a}}$ are the axial currents of the gauge field and
the Dirac particles,
\begin{align}
j^k_\text{\rm{a}} &= \partial^k_{\;\:l} A^l_\text{\rm{a}} - \Box A^k_\text{\rm{a}} \\
J^i_\text{\rm{a}} &= \sum_{k=1}^{n_f} \overline{\Psi_k} \gamma^5 \gamma^i \Psi_k
- \sum_{l=1}^{n_a} \overline{\Phi_l} \gamma^5 \gamma^i \Phi_l\:. \label{JDirac}
\end{align}
As in~\eqref{particles}, the wave functions~$\Psi_k$ and~$\Phi_l$ denote
the occupied particle and anti-particle states, respectively.
The constants~$C_0$ and~$C_2$ in~\eqref{daeq} are empirical parameters which
take into account the unknown microscopic structure of space-time. For a given regularization method,
these constants can be computed as functions of the fermion masses.

For clarity, we point out that the Dirac current~\eqref{JDirac} involves only the particle and
anti-particle states of the system, but not the states forming the Dirac sea.
The reason is that the contributions by the sea states cancel each other in
our action principle. As a consequence, only the deviations from the completely filled sea
configuration contribute to the Dirac current.
In the continuum limit, pair creation is described following Dirac's
original idea by removing a sea state and occupying instead a particle state.
To avoid confusion, we mention that the wave functions~$\Psi_k$ and~$\Phi_l$ need to be
suitably orthonormalized. Taking this into account, the sum of the one-particle currents in~\eqref{JDirac}
is indeed the same as the expectation value of the current operator computed for
the Hartree-Fock state obtained by taking the wedge product of the wave
functions~$\Psi_k$ and~$\Phi_l$.

We finally remark that more realistic models are obtained if one describes the vacuum
instead of~\eqref{seavac} by a direct sum of several sectors. The larger freedom in
perturbing the resulting Dirac operator gives rise to several effective gauge fields, which couple to the
fermions in a specific way. As shown in~\cite[Chapters~6-8]{PFP}, this makes it possible to realize
the gauge groups and couplings of the standard model.
The derivation of the corresponding field equations is work in progress.

\section{A New Mechanism for the Generation of Boson Masses}
The term~$C_2\, A^k_\text{\rm{a}}$ in~\eqref{daeq} gives the axial field a rest mass~$M = 
\sqrt{C_2/C_0}$. This bosonic mass term is surprising, because in standard
gauge theories a boson can be given a mass only by the Higgs mechanism of spontaneous
symmetry breaking. We now explain how the appearance of the mass term in~\eqref{daeq}
can be understood on a non-technical level (for more details see~\cite[\S6.2 and~\S8.5]{sector}).

In order to see the connection to gauge theories, it is helpful to consider the behavior of the Dirac operator
and the fermionic projector under gauge transformations. We begin with the familiar gauge
transformations of electrodynamics, for simplicity in the case~$m=0$ of massless
fermions. Thus assume that we have a pure gauge potential $A = \partial \Lambda$ with
a real function~$\Lambda(x)$. This potential can be inserted into the Dirac operator by the
transformation
\[ i \Pdd \;\rightarrow\; e^{i \Lambda(x)} i \Pdd\, e^{-i \Lambda(x)}
= i \Pdd + (\Pdd \Lambda) \:, \]
showing that the electromagnetic potential simply describes the phase transformation
$\Psi(x) \rightarrow e^{i \Lambda(x)} \Psi(x)$ of the wave functions.
Since the multiplication operator~$U=e^{i \Lambda}$ is unitary with respect to the
inner product~\eqref{stip}, it preserves the normalization of the fermionic states.
Thus in view of~\eqref{Pkernel}, the kernel of the fermionic projector simply transforms
according to
\[ P(x,y) \;\rightarrow\; e^{i \Lambda(x)} P(x,y)\, e^{-i \Lambda(y)} \:. \]
When forming the closed chain~\eqref{cchain}, the phase factors drop out.
This shows that our action principle is {\em{gauge invariant}} under the local
$U(1)$-transformations of electrodynamics.

We next consider an axial potential~$A_\text{a}$ as appearing in~\eqref{daeq}.
A pure gauge potential~$A_\text{a}= \partial \Lambda$ can be generated by the transformation
\[ i \Pdd \;\rightarrow\; e^{i \gamma^5 \Lambda(x)} i \Pdd\, e^{i \gamma^5 \Lambda(x)}
= i \Pdd + \gamma^5 (\Pdd \Lambda) \:, \]
suggesting that the kernel of the fermionic projector should be transformed according to
\[ P(x,y) \;\rightarrow\; e^{-i \gamma^5 \Lambda(x)} P(x,y)\, e^{-i \gamma^5 \Lambda(x)} \:. \]
The main difference compared to the electromagnetic case is that now the
transformation operator~$U=e^{-i \gamma^5 \Lambda(x)}$ is {\em{not}} unitary with
respect to the inner product~\eqref{stip}.
This leads to the technical complication that we need to be concerned about the normalization
of the fermionic states. More importantly, the phases no longer drop out of the closed chain, because
\begin{align*}
A_{xy} \rightarrow & \left( e^{-i \gamma^5 \Lambda(x)} P(x,y)\, e^{-i \gamma^5 \Lambda(x)} \right)
\left( e^{-i \gamma^5 \Lambda(y)} P(y,x)\, e^{-i \gamma^5 \Lambda(x)} \right) \\
&= e^{-i \gamma^5 \Lambda(x)} P(x,y)\, e^{-2 i \gamma^5 \Lambda(y)} P(y,x)\,
e^{-i \gamma^5 \Lambda(x)} \:.
\end{align*}
This shows that in general, our action is not invariant under axial gauge transformations.
As a consequence, the appearance of the axial potential in the field equations does not
contradict gauge invariance.

A more detailed analysis shows that the above axial transformation indeed changes only the
phases of the eigenvalues~$\lambda_i$ of the closed chain, and these phases
drop out when taking their absolute values as appearing in the closed chain.
But repeating the above argument in the case~$m>0$ of massive fermions, one finds additional
contributions proportional to~$m^2 A_a$ which even affect the absolute values~$|\lambda_i|$.
These contributions are responsible for the bosonic mass term in the field equations.

In simple terms, the bosonic mass arises because the corresponding potential does not describe
a local symmetry of our system. More specifically, an axial gauge transformation
changes the relative phase of the left- and right-handed components of the fermionic projector.
This relative phase does change the physical system and is thus allowed to enter the physical equations.
In order to get a closer connection to the Higgs mechanism, one can say that the axial gauge
symmetry is spontaneously broken by the states of the Dirac sea,
because they distinguish the relative phase of the left- and right-handed components of the
fermionic projector.

\section{The Vacuum Polarization} \label{secpolarize}
\begin{figure}
\includegraphics{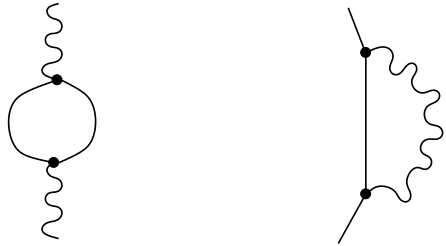}%
\caption{A fermionic loop diagram (left) and a bosonic loop diagram (right).}
\label{figfeynman}
\end{figure}

We now describe how the one-loop vacuum polarization arises in the fermionic projector approach
and compare the situation with perturbative QFT. For the derivation of the field equations
in Section~\ref{secfield}, we considered the singular contribution~$P^{\text{sing}}(x,y)$ in~\eqref{Pdecomp}, but we disregarded the noncausal contribution~$P^\text{reg}$.
Analyzing the latter contribution in the continuum limit gives rise to correction terms
to the field equations~\eqref{daeq} of the form
\beq \label{correction}
- f_{[0]}* j^k_\text{\rm{a}} + 6 f_{[2]}* A^k_\text{\rm{a}} \:,
\eeq
where~$f_{[p]}$ are explicit Lorentz invariant distributions and the star denotes convolution
(see~\cite[Theorem~8.2]{sector}). These corrections can already be understood in
Dirac's decomposition~\eqref{Rab} as the ``polarization effect'' as described
by the regular function~$R_b$.
In the static situation, the term~$- f_{[0]}* j^k_\text{\rm{a}}$ reduces to the axial analogue
of the well-known Uehling potential~\cite{uehling} (see~\cite[\S8.2]{sector}),
whereas the term~$6 f_{[2]}* A^k_\text{\rm{a}}$ can be regarded as a correction to the bosonic
mass term. We have thus reproduced the standard vacuum polarization, which is described in more
modern language by the Feynman diagram involving one fermion loop in Figure~\ref{figfeynman} (left).

The connection to the Uehling correction in standard QFT can be understood
most easily by going back to the original papers~\cite{dirac3, heisenberg2, uehling}.
Heisenberg starts from Dirac's decomposition~\eqref{Rab}. Motivated by symmetry considerations
and physical arguments, he gives a procedure for disregarding the singularities, so that only a
regular contribution remains.
This regular contribution gives rise to the Uehling potential.
Similarly, the starting point in~\cite{sector} is the decomposition of the
fermionic projector~\eqref{Pdecomp}. The main difference is that now the singular terms
are not disregarded or removed, but they are carried along all the way.
However, the singular terms drop out of the Euler-Lagrange equations corresponding to our
action principle~\eqref{actprinciple}. In this way, all divergences disappear.
The remaining finite contributions to~$P^\text{sing}$ give rise to
the bosonic current and mass terms in the resulting field equations~\eqref{daeq},
whereas~$P^\text{reg}$ describes the vacuum polarization.
The main advantage of the fermionic projector approach is that no counter terms are
needed. The back-reaction of the Dirac sea on the electromagnetic field is finite,
no divergences occur. Moreover, as we do not need counter terms, the
setting immediately becomes background independent. It is to be expected (although it has not yet
been worked out in detail) that the singularities of the fermionic projector will also drop out of the Euler-Lagrange equations if one sets up the theory in curved space-time.

In modern QFT, the vacuum polarization is still described as in the original
papers, with the only difference
that the singularities are now removed more systematically by a normal ordering of the field operators.
In the interacting situation, the subtle point is to choose the correct ``dressing'' of the electrons.
This means that one must distinguish a subspace of the Fock space as describing the Dirac sea;
then the normal ordering is performed with respect to this subspace.
In~\cite{bach} a quantized Dirac field is considered which interacts
with a Coulomb field and a magnetic field. It is shown that the resulting Hamiltonian
is positive, provided that the atomic numbers and the fine structure
constant are not too big. However, the chosen dressing has the shortcoming
that polarization effects are suppressed. A more careful
analysis is given in the series of papers~\cite{hainzl+sere2, hainzl+sere1,
hainzl+solovej2, gravejat+lewin+sere}, where the vacuum state is constructed
for a system of Dirac particles with electrostatic interaction in the Bogoliubov-Dirac-Fock approximation,
and the question of renormalization is addressed.
The conclusion of this analysis is that for mathematical consistency, one must take into account all the
states forming the Dirac sea. Furthermore, the interaction ``mixes'' the states in such a way that
it becomes impossible to distinguish between the particle states and the states of the Dirac sea.
Thus, despite the use of a very different mathematical framework, the physical picture in these papers
is quite similar to that of the fermionic projector approach.

\section{General Loop Diagrams}
So far, we only considered a Feynman diagram involving a
fermion loop. Let us now consider how to obtain Feynman diagrams which involve
bosonic loops: In the continuum limit, the system is described by the
partial differential equations~\eqref{daeq}. Here the bosonic potential~$A_\text{\rm{a}}$
is not quantized; it is simply a classical field. But the system~\eqref{daeq} is nonlinear,
and as shown in~\cite[\S8.4]{sector}, treating this nonlinearity perturbatively gives rise to
the bosonic loop diagram in Figure~\ref{figfeynman} (right), as well as higher order bosonic
loop diagrams. Taking the corrections~\eqref{correction} into account, one also gets the
diagrams with fermion loops. In this way, one gets all the usual Feynman diagrams.
But there are also differences. Since the analysis of the diagrams has not yet been carried
out systematically, we merely state the potential effects as open problems:
\begin{itemize}
\item It is not clear whether the usual divergences of the bosonic loop diagram
in Figure~\ref{figfeynman} (right) can be associated with a singularity of the fermionic projector
which drops out of our action principle (similar to the explanation for the fermionic loop diagram
in Section~\ref{secpolarize}). More generally, it is an open problem whether the
bosonic loop diagrams necessary diverge. In particular, it seems promising to try to
avoid the divergences completely by a suitable choice of the bosonic Green's function.
This analysis might reveal a connection to the ``causal approach'' by
Epstein and Glaser~\cite{epstein+glaser} and Scharf~\cite{scharf}.
\item The main difference of the perturbation expansion in the fermionic projector approach
is that instead of working with the Feynman propagator, the normalization conditions for the
sea states enforce a non-trivial combinatorics of operator products involving different types
of Green's functions and fundamental solutions (for details see~\cite{grotz}).
This difference has no influence on the singularities of the resulting Feynman diagrams, 
and thus we expect that the renormalizability of the theory is not affected.
But the higher-loop radiative corrections should depend on the detailed combinatorics, giving the
hope to obtain small deviations from standard QFT which might be tested experimentally.
\end{itemize}

\section{Violation of Causality}
As explained in Section~\ref{secpolarize}, the correction terms in~\eqref{correction}
can also be understood in the framework of standard QFT via fermionic loop diagrams (like in
Figure~\ref{figfeynman} (left)). However, the detailed analysis
of the correction terms in position space as carried out in~\cite[Chapter~8 and Appendix~D]{sector}
reveals an underlying structure which is not apparent in the usual description in momentum space.
Namely, the correction term~\eqref{correction} 
violates causality in the sense that the future can influence the past!
To higher order in the bosonic potential, even space-time points
with spacelike separation can influence each other.
At first sight, a violation of causality seems worrisome because it contradicts experience
and seems to imply logical inconsistencies. However, these non-causal correction
terms are only apparent on the Compton scale, and furthermore they
are too small for giving obvious contradictions to physical observations.
But they might open the possibility for future experimental tests.
For a detailed discussion of the causality violation we refer to~\cite[\S8.2 and~\S8.3]{sector}.

In order to understand how the violation of causality comes about, it is helpful to briefly discuss
the general role of causality in the fermionic projector approach. We first point out that in
discrete space-time, causality does not arise on the fundamental level.
But for a given minimizer of our action principle, Definition~\ref{defcausal} gives us the notion of
a ``discrete causal structure.'' This notion is compatible with our action principle in the
sense that space-time points~$x$ and~$y$ with spacelike separation do not influence each
other via the Euler-Lagrange equations. This can be seen as follows:
According to our definition, for such space-time points the eigenvalues of the closed chain
all have the same absolute value. Using the specific form of the Lagrangian~\eqref{Ldef},
this implies that the Lagangian and its first variation vanish. This in turn implies
that~$A_{xy}$ drops out of the Euler-Lagrange equations.
We conclude that our action principle is ``causal'' in the sense that no spacelike
influences are possible. But at this stage, no time direction is distinguished, and therefore
there is no reason why the future should not influence the past.

The system of hyperbolic equations~\eqref{daeq} obtained in the continuum limit
is causal in the sense that given initial data has a unique time evolution.
Moreover, we have finite propagation speed, meaning that no information can travel faster than the
speed of light. Thus in the continuum limit we recover the usual notion of causality. However, the fermionic
projector~$P^\text{sea}$ is not defined via an initial value problem, but it is
a global object in space-time (see~\cite[Chapter~2]{PFP}).
As a consequence, the contribution~$P^\text{reg}$
in~\eqref{Pdecomp} is noncausal in the sense that the future influences the past.
Moreover, to higher order in the bosonic potential
the normalization conditions for the fermions give rise to nonlocal constraints.
As a consequence, the bosonic potential may influence $P(x,y)$ even for spacelike
distances.

\section{Entanglement and Second Quantization}
Taking the wedge product of the one-particle wave functions,
\[ \Psi_1 \wedge \cdots \wedge \Psi_f \:, \]
and considering the continuum limit, we obtain a system of classical bosonic fields
coupled to a fermionic Hartree-Fock state. Although this setting gives rise to the Feynman
diagrams, it is too restrictive for describing all quantum effects observed in nature.
However, as shown in~\cite{entangle}, the framework of the fermionic projector
also allows for the description of general second quantized fermionic and bosonic fields.
In particular, it is possible to describe entanglement.

The derivation of these results is based on the assumption that space-time should
have a non-trivial microstructure. In view of our concept of discrete space-time,
this assumption seems natural. Homogenizing the microstructure, one obtains an
effective description of the system by a vector in  the fermionic or bosonic Fock space.
This concept, referred to as the {\em{microscopic mixing of decoherent subsystems}},
is worked out in detail in~\cite{entangle}. In~\cite{dice2010}, the methods and results are
discussed with regard to decoherence phenomena and the measurement problem.

\section{Conclusions and Outlook}
Combining our results, we obtain a formulation of QFT which
is consistent with perturbative QFT but has surprising additional features.
First, we find a new mechanism for the generation of masses of gauge bosons
and obtain new types of corrections to the field equations which violate causality.
Moreover, our model involves fewer free parameters, and the structure of
the interaction is completely determined by our action principle.
Before one can think of experimental tests, one clearly needs to work out a more realistic model
which involves all elementary particles and includes all interactions observed in nature.
As shown in~\cite[Chapters~6--8]{PFP}, a model involving 24
Dirac seas is promising because the resulting gauge fields have striking similarity to the
standard model. Furthermore, the underlying diffeomorphism invariance gives agreement with
the equivalence principle of general relativity.
Thus working out the continuum limit of this model in detail will lead to a 
formulation of QFT which is satisfying conceptually and makes
quantitative predictions to be tested in future experiments.

\Thanks{{\em{Acknowledgments:}} I would like to thank Bertfried Fauser, Christian Hainzl and the
referees for helpful comments on the manuscript.}


\begin{thebibliography}{10}

\bibitem{bach}
V.~Bach, J.-M. Barbaroux, B.~Helffer, and H.~Siedentop, \emph{On the stability
  of the relativistic electron-positron field}, Comm. Math. Phys. \textbf{201}
  (1999), no.~2, 445--460.

\bibitem{baer+fredenhagen}
C~B\"ar and K.~Fredenhagen~(eds), \emph{Quantum field theory on curved
  spacetimes}, Lecture Notes in Physics, vol. 786, Springer Verlag, Berlin,
  2009.

\bibitem{christensen}
S.M. Christensen, \emph{Vacuum expectation value of the stress tensor in an
  arbitrary curved background: the covariant point-separation method}, Phys.
  Rev. D (3) \textbf{14} (1976), no.~10, 2490--2501.

\bibitem{collins}
J.C. Collins, \emph{Renormalization}, Cambridge Monographs on Mathematical
  Physics, Cambridge University Press, Cambridge, 1984.

\bibitem{merkl}
D.-A. Deckert, D.~D\"urr, F.~Merkl, and M.~Schottenloher, \emph{Time evolution
  of the external field problem in {QED}}, arXiv:0906.0046 [math-ph] (2009).

\bibitem{dirac2}
P.A.M. Dirac, \emph{A theory of electrons and protons}, Proc. R. Soc. Lond. A
  \textbf{126} (1930), 360--365.

\bibitem{dirac3}
\bysame, \emph{Discussion of the infinite distribution of electrons in the
  theory of the positron}, Proc. Camb. Philos. Soc. \textbf{30} (1934),
  150--163.

\bibitem{dirac4}
\bysame, \emph{Directions in physics}, Wiley-Interscience [John Wiley \& Sons],
  New York, 1978, Five lectures delivered during a visit to Australia and New
  Zealand, August--September, 1975.

\bibitem{heisenberg}
H.-P. D{\"u}rr, W.~Heisenberg, H.~Mitter, S.~Schlieder, and K.~Yamazaki,
  \emph{Zur {T}heorie der {E}lementarteilchen}, Z. Naturf. \textbf{14a} (1959),
  441--485.

\bibitem{dyson2}
F.J. Dyson, \emph{The {$S$} matrix in quantum electrodynamics}, Phys. Rev.
  \textbf{75} (1949), 1736--1755.

\bibitem{epstein+glaser}
H.~Epstein and V.~Glaser, \emph{The role of locality in perturbation theory},
  Ann. Inst. H. Poincar\'e Sect. A (N.S.) \textbf{19} (1973), 211--295.

\bibitem{feynman}
R.~Feynman, \emph{The theory of positrons}, Phys. Rev. \textbf{76} (1949),
  749--759.

\bibitem{fierz+scharf}
H.~Fierz and G.~Scharf, \emph{Particle interpretation for external field
  problems in {QED}}, Helv. Phys. Acta \textbf{52} (1979), no.~4, 437--453.

\bibitem{PFP}
F.~Finster, \emph{The principle of the fermionic projector}, hep-th/0001048,
  hep-th/0202059, hep-th/0210121, AMS/IP Studies in Advanced Mathematics,
  vol.~35, American Mathematical Society, Providence, RI, 2006.

\bibitem{sector}
\bysame, \emph{An action principle for an interacting fermion system and its
  analysis in the continuum limit}, arXiv:0908.1542 [math-ph] (2009).

\bibitem{lrev}
\bysame, \emph{From discrete space-time to {M}inkowski space: Basic mechanisms,
  methods and perspectives}, arXiv:0712.0685 [math-ph], Quantum Field Theory
  (B.~Fauser, J.~Tolksdorf, and E.~Zeidler, eds.), Birkh\"auser Verlag, 2009,
  pp.~235--259.

\bibitem{entangle}
\bysame, \emph{Entanglement and second quantization in the framework of the
  fermionic projector}, arXiv:0911.0076 [math-ph], J. Phys. A: Math. Theor.
  \textbf{43} (2010), 395302.

\bibitem{dice2010}
\bysame, \emph{The fermionic projector, entanglement, and the collapse of the
  wave function}, arXiv:1011.2162 [quant-ph], to appear in the Proceedings of
  DICE2010 (2011).

\bibitem{grotz}
F.~Finster and A.~Grotz, \emph{The causal perturbation expansion revisited:
  Rescaling the interacting {D}irac sea}, arXiv:0901.0334 [math-ph], J. Math.
  Phys. \textbf{51} (2010), 072301.

\bibitem{fulling+sweeny+wald}
S.A. Fulling, M.~Sweeny, and R.M. Wald, \emph{Singularity structure of the
  two-point function quantum field theory in curved spacetime}, Comm. Math.
  Phys. \textbf{63} (1978), no.~3, 257--264.

\bibitem{glimm+jaffe}
J.~Glimm and A.~Jaffe, \emph{Quantum physics, a functional integral point of
  view}, second ed., Springer-Verlag, New York, 1987.

\bibitem{gravejat+lewin+sere}
P.~Gravejat, M.~Lewin, and E.~S{\'e}r{\'e}, \emph{Renormalization and
  asymptotic expansion of {D}irac's polarized vacuum}, arXiv:1004.1734v1
  (2010).

\bibitem{hainzl+sere2}
C.~Hainzl, M.~Lewin, and E.~S{\'e}r{\'e}, \emph{Existence of a stable polarized
  vacuum in the {B}ogoliubov-{D}irac-{F}ock approximation},
  arXiv:math-ph/0403005, Comm. Math. Phys. \textbf{257} (2005), no.~3,
  515--562.

\bibitem{hainzl+sere1}
\bysame, \emph{Self-consistent solution for the polarized vacuum in a no-photon
  {QED} model}, arXiv:physics/0404047, J. Phys. A: Math. Theor. \textbf{38}
  (2005), no.~20, 4483--4499.

\bibitem{hainzl+solovej2}
C.~Hainzl, M.~Lewin, E.~S{\'e}r{\'e}, and J.P. Solovej, \emph{A minimization
  method for relativistic electrons in a mean-field approximation of quantum
  electrodynamics}, arXiv:0706.1486 [physics.atom-ph], Phys. Rev. A \textbf{76}
  (2007), 052104.

\bibitem{heisenberg2}
W.~Heisenberg, \emph{Bemerkungen zur {D}iracschen {T}heorie des {P}ositrons},
  Z. Phys. \textbf{90} (1934), 209--231.

\bibitem{klaus}
M.~Klaus, \emph{Nonregularity of the {C}oulomb potential in quantum
  electrodynamics}, Helv. Phys. Acta \textbf{53} (1980), no.~1, 36--39.

\bibitem{klaus+scharf1}
M.~Klaus and G.~Scharf, \emph{The regular external field problem in quantum
  electrodynamics}, Helv Phys. Acta \textbf{50} (1977), no.~6, 779--802.

\bibitem{klaus+scharf2}
\bysame, \emph{Vacuum polarization in {F}ock space}, Helv. Phys. Acta
  \textbf{50} (1977), no.~6, 803--814.

\bibitem{nenciu+scharf}
G.~Nenciu and G.~Scharf, \emph{On regular external fields in quantum
  electrodynamics}, Helv. Phys. Acta \textbf{51} (1978), no.~3, 412--424.

\bibitem{peskin+schroeder}
M.E. Peskin and D.V. Schroeder, \emph{An introduction to quantum field theory},
  Addison-Wesley Publishing Company Advanced Book Program, Reading, MA, 1995.

\bibitem{radzikowski}
M.J. Radzikowski, \emph{Micro-local approach to the {H}adamard condition in
  quantum field theory on curved space-time}, Comm. Math. Phys. \textbf{179}
  (1996), no.~3, 529--553.

\bibitem{scharf}
G.~Scharf, \emph{Finite Quantum Electrodynamics}, Texts and Monographs in
  Physics, Springer-Verlag, Berlin, 1989.

\bibitem{schwinger}
J.~Schwinger, \emph{Quantum electrodynamics. {I}. {A} covariant formulation},
  Phys. Rev. \textbf{74} (1948), 1439--1461.

\bibitem{serber}
R.~Serber, \emph{Linear modifications of the {M}axwell field equations}, Phys.
  Rev. \textbf{48} (1935), 49--54.

\bibitem{uehling}
E.A. Uehling, \emph{Polarization effects in the positron theory}, Phys. Rev.
  \textbf{48} (1935), 55--63.

\end{thebibliography}
\def\dbar{\leavevmode\hbox to 0pt{\hskip.2ex \accent"16\hss}d}
\providecommand{\bysame}{\leavevmode\hbox to3em{\hrulefill}\thinspace}
\providecommand{\MR}{\relax\ifhmode\unskip\space\fi MR }
\providecommand{\MRhref}[2]{%
  \href{http://www.ams.org/mathscinet-getitem?mr=#1}{#2}
}
\providecommand{\href}[2]{#2}

\end{document}